\documentclass[10pt,conference]{IEEEtran}
\usepackage{cite}
\usepackage{amsmath,amssymb,amsfonts}
\usepackage{algorithmic}
\usepackage{graphicx}
\usepackage{textcomp}
\usepackage{xcolor}
\usepackage{subcaption}
\usepackage{hyperref}
\usepackage{url}
\usepackage{booktabs}
\begin{document}
\title{Q\&A MAESTRO: Q\&A Post Recommendation for Fixing Java Runtime Exceptions}

\makeatletter
\newcommand{\linebreakand}{%
  \end{@IEEEauthorhalign}
  \hfill\mbox{}\par
  \mbox{}\hfill\begin{@IEEEauthorhalign}
}
\makeatother

\author{
  \IEEEauthorblockN{Yusuke Kimura}
  \IEEEauthorblockA{Fujitsu Limited\\
    Kawasaki, Japan\\
    yusuke-kimura@fujitsu.com}
  \and
  \IEEEauthorblockN{Takumi Akazaki}
  \IEEEauthorblockA{Fujitsu Limited\\
    Kawasaki, Japan\\
    akazaki.takumi@fujitsu.com}
  \and
  \IEEEauthorblockN{Shinji Kikuchi}
  \IEEEauthorblockA{Fujitsu Limited\\
    Kawasaki, Japan\\
    skikuchi@fujitsu.com}
  \linebreakand 
  \IEEEauthorblockN{Sonal Mahajan}
  \IEEEauthorblockA{Fujitsu Research of America, Inc.\\
    Sunnyvale, USA\\
    smahajan@fujitsu.com}
  \and
  \IEEEauthorblockN{Mukul R. Prasad}
  \IEEEauthorblockA{Fujitsu Research of America, Inc.\\
    Sunnyvale, USA\\
    mukul@fujitsu.com}
}

\maketitle

\begin{abstract}
Programmers often use Q\&A sites (e.g., Stack Overflow) to understand a root cause of program bugs.
Runtime exceptions is one of such important class of bugs that is actively
discussed on Stack Overflow.
However, it may be difficult for beginner programmers to come up with appropriate keywords for search. Moreover, they need to switch their attentions between IDE and browser, and it is time-consuming. 
To overcome these difficulties, we proposed a method, ``Q\&A MAESTRO'', to find suitable Q\&A posts automatically for Java runtime exception by utilizing structure information of codes described in programming Q\&A website.
In this paper, we describe a usage scenario of IDE-plugin, the architecture and user interface of the implementation, and results of user studies.
A video is available at \url{https://youtu.be/4X24jJrMUVw}.
A demo software is available at \url{https://github.com/FujitsuLaboratories/Q-A-MAESTRO}.
\end{abstract}

\section{Introduction}
While developing software, debugging is an important process. 
According to a survey \cite{debug}, 50\% of the total time of programmers is spent for debugging.
In particular, runtime exceptions have a significant impact on system availability and crashes \cite{10.1145/1181309.1181314}.
As of March 2020 , Stack Overflow \cite{so} has more than 190,000 posts related to Java runtime exception, which is 7\% of Java posts.

A study \cite{7997917} collected mouse events to find out how programmers spend their time. It found that average 43 percent of the total time was used for searching solutions to fix bugs.
Therefore, in order to improve the productivity of programmers, we need to help them do those time-consuming tasks quickly.

In our research paper \cite{maestro}, we proposed a method ``Q\&A MAESTRO'' to automatically suggest Q\&A posts for Java runtime exceptions. (Note: In this paper, we have changed the name of the technique proposed in the research paper \cite{maestro} from ``MAESTRO" to ``Q\&A MAESTRO"  due to the trademark reason.)
It has three features as follows. It achieved better suggestion quality compared to other studies.
\begin{itemize}
  \item Using question code snippets and comparing it to buggy user code to decide the relevancy of the post
  \item Proposing ``Failure scenario localization (FSL)", in which answer code snippets are used to identify the buggy lines in question code snippets, and ignore other lines
  \item Suggesting a new data structure: APG (Abstract Program Graph), which is specific for Q\&A search
\end{itemize}

This paper introduces the implementation of Q\&A MAESTRO. The main contribution is as follows.
\begin{itemize}
  \item Implementing IDE plugin so that users can solve problems without leaving IDE to Web browser for search
  \item Designing and implementing client-server architecture as in Fig. \ref{fig:arch} so that analysis and search tasks are executed on the server side
  \item Conducting user studies of the implementation and explaining feedbacks from the trial users
\end{itemize}

\begin{figure}[tb]
  \includegraphics[width=\linewidth]{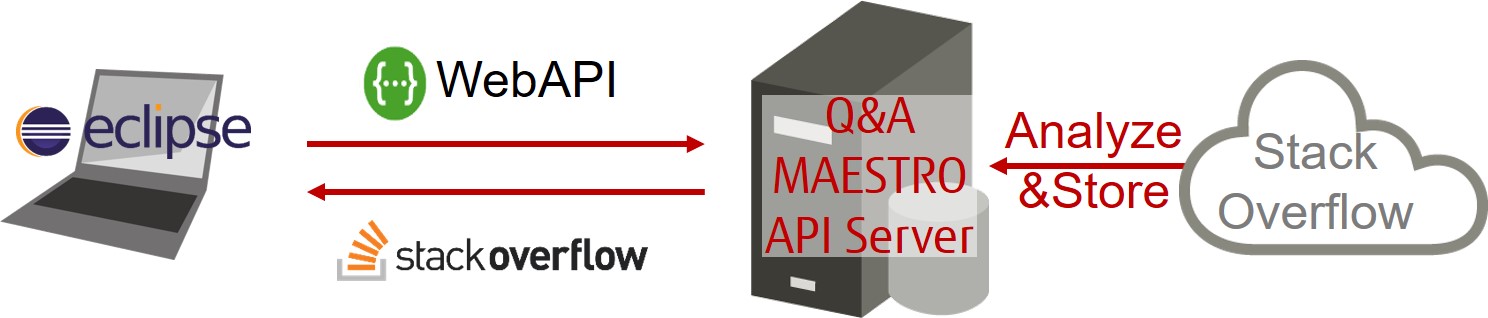}
  \caption{Server-client architecture}
  \label{fig:arch}
\end{figure}

The organization of this paper is as follows. Section \ref{sec:2} introduces researches related to Q\&A post suggestions. Section \ref{sec:3} summarizes our proposed method in our research paper. Section \ref{sec:4} describes the usage scenario of Q\&A MAESTRO and its implementation.
Section \ref{sec:5} summarizes the experiments in our research paper at first, and then, describes the user studies in two different groups in our company. Section \ref{sec:6} concludes the paper.

\section{Related Work}\label{sec:2}
Several methods have been proposed for Q\&A post suggestions for bug fixing.
Prompter \cite{prompter} proposed a method to find Q\&A posts which are similar to the piece of code being edited in the IDE. This method relies on extracting important keywords for search, not taking into account the structural information of source codes in the post.
FaCoY \cite{facoy} also proposed a method to search code in Q\&A articles similar to the given  code snippet. It still focuses on the natural language description in question posts and corresponding answer posts.
Q\&A MAESTRO is different from the above studies because we utilize structural information in the codes in Q\&A posts.

Deckard \cite{deckard} proposed a code clone detection method for finding similar codes.
It finds not only exactly the same codes, but also similar codes.
It uses AST (Abstract Syntax Tree) for analyzing codes, but our method uses APG (Abstract Program Graph) generated by abstracting AST to realize more accurate suggestions.

\section{Q\&A post suggestion technique}\label{sec:3}
This section summarizes our proposed method in our research paper \cite{maestro}.

The method can be divided into two phases: offline mining/analysis and real-time matching.
The overall flow of the method is shown in Fig. \ref{fig:flow}.
In both phases, any codes are converted to tree structures for analysis. While AST is common  for program analysis, our method uses APG generated by abstracting AST.

\begin{figure}[tb]
  \includegraphics[bb=0 0 735 309, width=\linewidth]{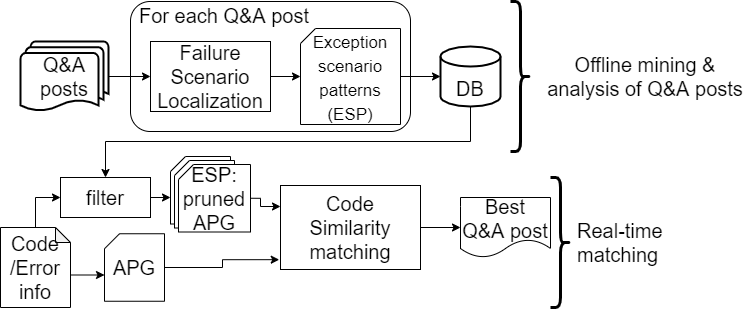}
  \caption{Overall Flow}
  \label{fig:flow}
\end{figure}

\subsection{APG (Abstract Program Graph)}
\subsubsection{About APG}
Many existing studies related to software engineering use AST to analyze codes. 
However, when finding similar code from Q\&A site, there is no need to focus too much on grammatical differences. Therefore, we consider that more abstracted structure is more suitable for Q\&A suggestion.

APG is designed to satisfy the above-mentioned properties. 
It is generated from AST through the following processes.
\begin{itemize}
  \item Merge neighbor nodes (e.g., function calls) into a simpler single node
  \item Add type information to a variable without it (e.g., if ``x=1'', then x is regarded as integer.)
  \item Convert semantically similar node types to the same node type (e.g., ``for'', ``while'' and ``for-each'' can be ``loop'')
  \item Change user-defined class name to general name like ``userClass1''
\end{itemize}

Please also see the illustrated example of APG in our research paper \cite{maestro}.

\subsubsection{APG structural alignment}
The goal of this process is to find the structural correspondence between two APGs.

Since APG is a tree structure, the minimum tree edit distance can be calculated.
To modify one tree structure into another, you can consider four operations: INSERT, DELETE, UPDATE and MATCH. 
MATCH nodes are the corresponding ones in the two trees.
Each operation can have cost, for example, (INSERT, DELETE, UPDATE, MATCH) = (1, 1, 1, 0). Then, we can find a set of operations whose total cost is the smallest. Such operations are called the minimum tree edit distance.
You can find APG structural alignment based on MATCH nodes.

\subsubsection{Code similarity matching}
The goal of this process is to calculate similarity score between two APGs.
The score is the weighted sum of some sub-scores. One of the sub-scores is construct similarity, which is the number of MATCH nodes in APG structural alignment.
Please refer to our research paper for further details.

\subsection{Offline mining \& analysis of Q\&A posts}
In this phase, we collect buggy code snippets from Q\&A sites and store them in Exception scenario patterns (ESP, pruned APG) so that we can utilize the information in the recommendation phase.

In Q\&A site, users can post multiple answers to one question post. Especially in Q\&A site for developers, like Stack Overflow, programmers can include codes in both question and answer posts.
Questioners usually do not know where in their codes is buggy. Therefore, they tend to submit codes which are not related to an error.
If a search is performed on the entire code in the posts, ones unrelated to the error may be proposed.

To overcome this difficulty, we need to localize the buggy part of the code in a question post. We can achieve this by Failure Scenario Localization (FSL), which consists of three steps as follows.
The basic idea of FSL is to find similar but not identical parts in Q\&A codes, and such part in the question code should be buggy.

\begin{enumerate}
  \item Convert a question code and answer code into APGs: $APG_Q$ and $APG_A$ respectively
  \item Perform APG structural alignment between $APG_Q$ and $APG_A$
  \item Only edited parts are kept in $APG_Q$, and other nodes are removed
\end{enumerate}

We refer to such pruned APG as ESP. ESPs are stored in the database.

\subsection{Real-time matching}\label{sec:2nd}
This phase is performed when users find runtime exceptions in their code.
The inputs of this phase are as follows.
\begin{itemize}
  \item Code with error (e.g., \texttt{class CAL\{int calc() \{...\}\}})
  \item Line number in the code where the error occurred
  \item Exception name (e.g., ``ArithmeticException'')
\end{itemize}
Q\&A MAESTRO provides relevant Q\&A posts by performing code similarity matching between ESPs stored in the database and APG converted from the input code causing the runtime exception.

This phase performs the following process. First, we translate the function containing the erroneous line into APG: $APG_i$.
Next, we identify the ESPs corresponding to the same exception of the input. Then, we recommend some of these ESPs which have high similarity score with $APG_i$.

\section{Tool Description}\label{sec:4}
\subsection{Usage Scenario: IDE Integration} \label{sec:scenario}

\begin{figure}[tb]
  \includegraphics[bb=0 0 1737.242490 1143.909671, width=\linewidth]{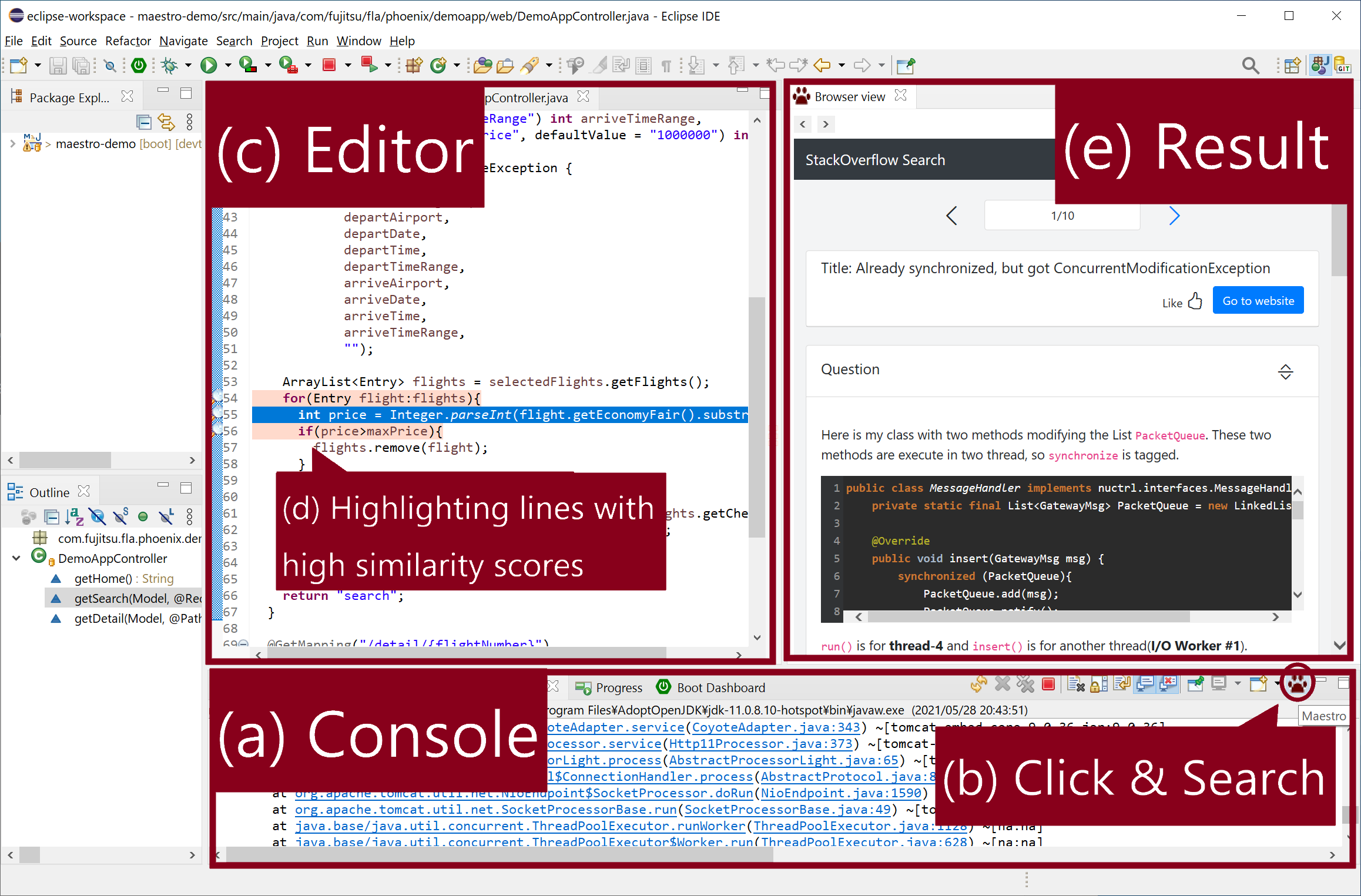}
  \caption{Eclipse plugin}
  \label{fig:eclipse}
\end{figure}

One usage scenario of Q\&A MAESTRO is IDE plugin.
The UI of the plugin is shown in Fig. \ref{fig:eclipse}.

When an error occurs, a console window on the bottom shows a stack trace of the error as (a) in Fig. \ref{fig:eclipse}.
You can get Q\&A posts automatically by just pressing the search button on the console menu (b).
The source code file containing the error is opened in the editor automatically (c), and the erroneous lines are highlighted (d).
The recommended Q\&A posts are provided as a web page and shown in a new IDE tab (e). 

The advantage of this usage scenario is that the search is highly automated.
In addition, the search results are also displayed in the IDE, so you can minimize context switching between IDE and Web browser for keyword search.

\subsection{Implementation}\label{sec:tool}
\subsubsection{Overall system architecture}
We developed a prototype software for the usage scenario mentioned above.

As in Fig. \ref{fig:arch}, it consists of two parts: server and IDE plugin. The prototype comes with an Eclipse plugin.
By dividing the system into server and client, clients do not need to hold the vast amount of data retrieved from Stack Overflow. Each IDEs needs each implementation, so the server provides a RESTful API for easy implementation.

In the server side, JavaParser \cite{javaparser} is used for analyzing Java code, and APTED \cite{apted} is used for calculating minimum tree edit distance. The indexed Q\&A posts are retrieved from Stack Overflow\cite{sodump}.

As alternative for users not using Eclipse, we also implemented stand-alone Web interface on the server.

\subsubsection{One click search}
To realize one click search, we derive the inputs described in Sec.\ref{sec:2nd} in the following way.

\begin{itemize}
\item Exception name: Since it is shown on the top of the stack trace, we derive it from there. Exception message is ignored.
\item Line number and file name: Since the stack trace shows us numerous file names and erroneous lines in its frames, we identify the proper ones from the first frame pointing to a user-owned code (not linked libraries), since we are not interested in identifying error in libraries developed by someone else. 
\end{itemize}

In case the above method does not find appropriate Q\&A posts, we prepared alternative search functions in which we can manually select a specific stack frame, or a line of code and exception name.

\subsubsection{Result UI}
The search results are shown in a summarized way as in Fig. \ref{fig:ui}, not directly showing original Stack Overflow web page. It shows a question post, one recommended answer post, and colored difference between the codes in the question and the answer. It can show us up to ten results so that users can determine the right answer from them.   

\begin{figure}[tb]
  \centering
  \includegraphics[width=0.9\linewidth, bb=0 0 1240.673177 1591.722178]{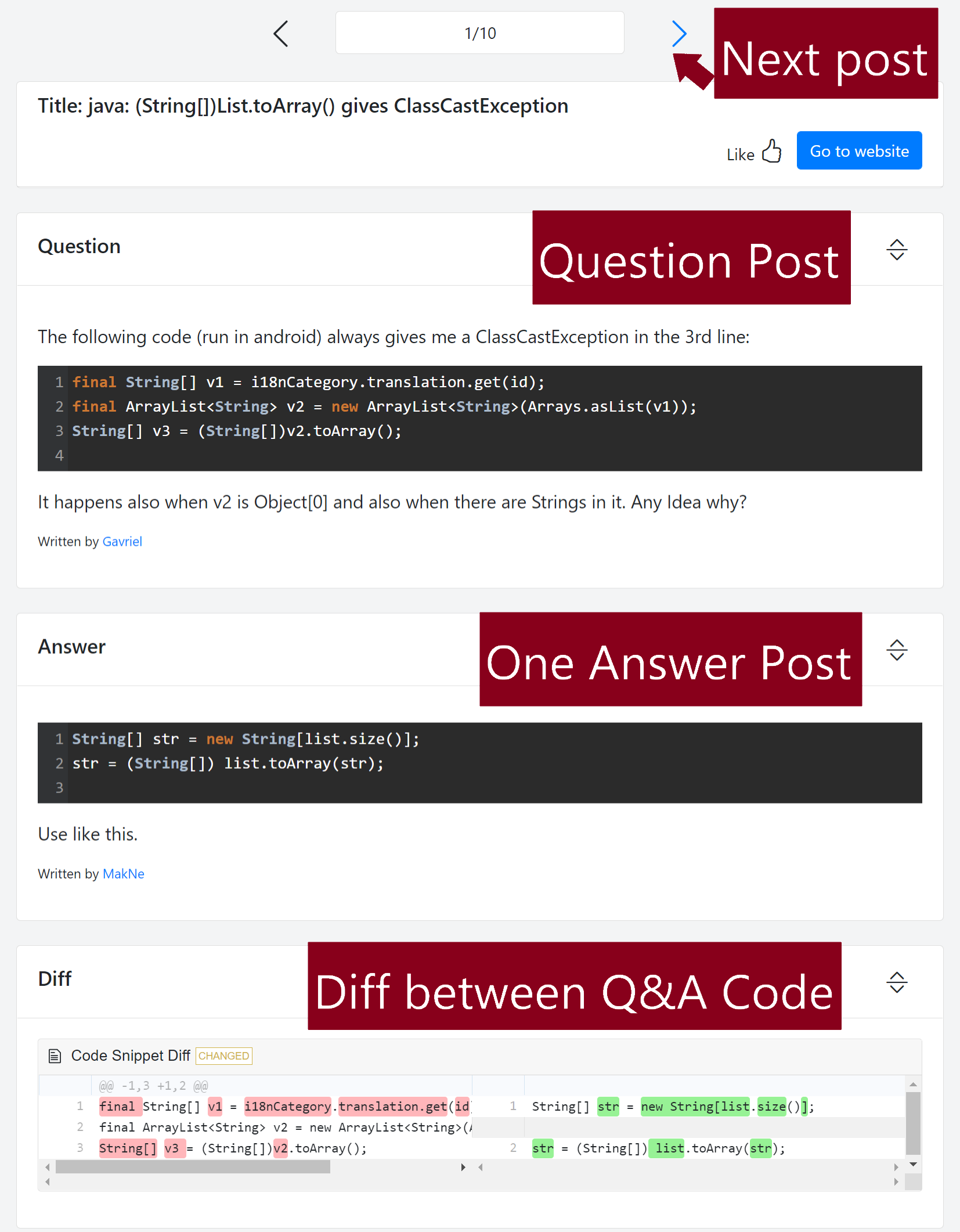}
  \caption{UI for a suggested post}
  \label{fig:ui}
\end{figure}

\section{Evaluation}\label{sec:5}
\subsection{Comparison with other techniques}
This section summarizes the experiments conducted in our research paper \cite{maestro}.
It compares Q\&A MAESTRO with three existing studies: Prompter\cite{prompter}, FaCoY\cite{facoy}, Deckard\cite{deckard}, and keyword search.
The suggested Q\&A posts are from Stack Overflow \cite{sodump}.
We used 78 open source codes in GitHub as a dataset \cite{dataset}, which causes runtime exceptions. Q\&A MAESTRO suggested useful articles for 70\% of the dataset, that was better than the others. You can find the exception names, the number of stored ESPs and the detailed results in the research paper.

The execution time of Q\&A MAESTRO in the above experiment was about 2.6 sec on median (average = 76 sec).
Please note that the implementation used in the research paper was a preliminary version. In the user studies, we implemented the system shown in \ref{sec:tool}, and its execution time was 0.2 sec on median (average = 0.45 sec). It is fast enough for interactive problems solving between users and Eclipse IDE.

\subsection{User Studies}
To survey the usability of the IDE plugin, we asked two groups in our company to use it in their own programming tasks for about two weeks.
We distributed the plugin to users, and they access to the central server we managed.
We asked them to complete an open-ended questionnaire on the three points: (1) Usability of the plugin (2) Accuracy of the suggested posts (3) Readability of the displayed results.

The first group is a team of ten programmers developing in-house systems. All of them are expert Java programmers and have several years of experience using Java in their work.
We received the feedbacks from them as shown in TABLE \ref{tab:experts}.

\begin{table}[ht]
  \caption{Feedbacks from experts}
  \label{tab:experts}
  \begin{tabular}{p{\linewidth}}\toprule
  Positive feedbacks \\\midrule
  * It is easy to use. \\
  * The suggestion are quite accurate.\\\toprule
  Negative feedbacks\\\midrule
  * For security reasons, there is some cases where we are not allowed to send our source code out of our department.\\
  * There is not much difference from searching by myself.\\
  * Just suggesting useful information is not enough. I prefer the code automatically repaired.\\\bottomrule  
  \end{tabular}
\end{table}

The second group is a Java programming training class for new employees. There are about 20 students, and most participants are Java or programming beginners.
We received feedbacks as in TABLE \ref{tab:beginners}.

\begin{table}[ht]
  \caption{Feedbacks from beginners}
  \label{tab:beginners}
  \begin{tabular}{p{\linewidth}}\toprule
  Positive feedbacks \\\midrule
  * It is easy to read results.\\
  * It is faster than searching by myself.\\\toprule
  Negative feedbacks\\\midrule
  * Some posts are not helpful for debugging. (Note: Especially for NullPointerException)\\
  * Some posts do not describe a root-cause of an error.\\\bottomrule
  \end{tabular}
\end{table}

Both groups were generally satisfied with the suggestion accuracy.
Q\&A MAESTRO mainly targets bugs due to lack of knowledge, so it worked properly for such bugs (e.g., lack of an argument for toArray function).
However, since NullPointerException can have variety of root causes, it is quite difficult to suggest appropriate posts for this exception.  
Moreover, Q\&A MAESTRO has difficulty in suggesting good posts for logical errors (e.g., ArrayIndexOutOfBounds, NegativeArraySize). 

The plugin was easy to use, so no one was confused about how to use it.
However, the expert programmers asked additional features, e.g., automatic bug-fix. 
On the other hand, the beginners were satisfied with the plugin's functionality, but requested more detailed Q\&A posts.

As stated above, both the experts and the beginners were generally satisfied with the plugin's usability and result UI. 
We also found that experts and beginners have different expectations for the plugin. Experts do not have problems with Q\&A post searching, and want advanced features such as automatic correction. While the IDE plugin contributed to shorten the time spent for debugging, most of them thought its effectiveness was marginal.
On the other hand, beginners have problems for finding and reading Q\&A posts, and actually realized the reduction of debugging time.
Therefore, the plugin is helpful for beginners especially. 

\section{Conclusion}\label{sec:6}
First, this paper explains our research paper\cite{maestro} which proposed a method ``Q\&A MAESTRO'' to automatically find relevant Q\&A posts to a Java runtime exception. 
Next, we propose IDE plugin as a use case of Q\&A MAESTRO and describe the software architecture and its implementation.
Finally, we explain the user studies in two groups, and introduce feedbacks from them. We find that our tool is helpful especially for beginners.

There are several challenges to make Q\&A MAESTRO more helpful.
The desirable Q\&A post should include both a cause of an error and how-to-fix.
Because Q\&A MAESTRO recognizes the differences in Q\&A codes, it has potential to be able to suggest posts including how-to-fix.
On the other hand, the cause of the error is often explained in natural language, and it cannot be derived by code structure analysis.
For a future work, we are planning to enhance Q\&A Maestro by adding features such as automated fix.

Moreover, Q\&A MAESTRO just suggests Q\&A posts, and the experts requested additional features like auto fix. This can be a future work.

\bibliographystyle{IEEEtran}
\bibliography{ase21}

\end{document}